\documentclass[twocolumn,showpacs,aps,prd,superscriptaddress,floatfix]{revtex4-2}
\usepackage{graphicx}  
\usepackage{dcolumn}   
\usepackage{bm}        
\usepackage{amssymb}   
\usepackage{amsfonts}  
\usepackage{hyperref}
\usepackage{amsmath}
\usepackage{amsthm}
\usepackage{mathtools}
\usepackage{xspace}
\usepackage{gensymb}
\usepackage{mathrsfs}

\def \gev  {\mbox{GeV}~}
\def \gevc {\mbox{GeV/$c$}~}
\def \gevcc{\mbox{GeV/$c^2$}~}

\def \bes  {\mbox{BESIII}~}

\def \gev  {\mbox{GeV}}
\def \gevc {\mbox{GeV/$c$}}
\def \gevcc{\mbox{GeV/$c^2$}}

\def\simge{\mathrel{
   \rlap{\raise 0.511ex \hbox{$>$}}{\lower 0.511ex \hbox{$\sim$}}}}
\def\simle{\mathrel{
   \rlap{\raise 0.511ex \hbox{$<$}}{\lower 0.511ex \hbox{$\sim$}}}}

\begin{document}
\title{\bf Study of the electromagnetic Dalitz decay \boldmath{$J/\psi \to e^+e^- \pi^0$}}

\author{
  \begin{small}
    \begin{center}
 M.~Ablikim$^{1}$, M.~N.~Achasov$^{4,c}$, P.~Adlarson$^{76}$, O.~Afedulidis$^{3}$, X.~C.~Ai$^{81}$, R.~Aliberti$^{35}$, A.~Amoroso$^{75A,75C}$, Q.~An$^{72,58,a}$, Y.~Bai$^{57}$, O.~Bakina$^{36}$
% ...
         \\
      \vspace{0.2cm}
      (BESIII Collaboration)\\
      \vspace{0.2cm} {\it
$^{1}$ Institute of High Energy Physics, Beijing 100049, People's Republic of China\\
% ...
$^{3}$ Bochum  Ruhr-University, D-44780 Bochum, Germany\\
$^{4}$ Budker Institute of Nuclear Physics SB RAS (BINP), Novosibirsk 630090, Russia\\
% ...
$^{35}$ Johannes Gutenberg University of Mainz, Johann-Joachim-Becher-Weg 45, D-55099 Mainz, Germany\\
$^{36}$ Joint Institute for Nuclear Research, 141980 Dubna, Moscow region, Russia\\
% ...
$^{57}$ Southeast University, Nanjing 211100, People's Republic of China\\
$^{58}$ State Key Laboratory of Particle Detection and Electronics, Beijing 100049, Hefei 230026, People's Republic of China\\
% ...
$^{72}$ University of Science and Technology of China, Hefei 230026, People's Republic of China\\
% ...
$^{75}$ University of Turin and INFN, (A)University of Turin, I-10125, Turin, Italy; (B)University of Eastern Piedmont, I-15121, Alessandria, Italy; (C)INFN, I-10125, Turin, Italy\\
$^{76}$ Uppsala University, Box 516, SE-75120 Uppsala, Sweden\\
% ...
$^{81}$ Zhengzhou University, Zhengzhou 450001, People's Republic of China\\
\vspace{0.2cm}
$^{a}$ Deceased\\
% ...
$^{c}$ Also at the Novosibirsk State University, Novosibirsk, 630090, Russia\\
% ...
      }\end{center}
    \vspace{0.4cm}
  \end{small}
}

\begin{abstract}
We report the first measurement of the di-electron invariant mass dependent transition form factor in the electromagnetic Dalitz decay $J/\psi \to e^+e^- \pi^0$ using  $(10087  \pm 44) \times 10^6$  $J/\psi$ events collected by the \bes detector.  A clear $\rho-\omega$ interference structure is observed, consistent with the pion form factor, which offers a novel approach to extract the hadronic vacuum polarization contribution to the anomalous muon magnetic moment ($a_{\mu}$) and refine the predictions of the Vector Meson Dominance (VMD) model and hadronic light-by-light contribution to $a_{\mu}$. By taking into account the contribution of this $\rho-\omega$ interference structure, the branching fraction of $J/\psi \to e^+e^- \pi^0$ in the full $e^+e^-$ invariant mass range is also measured  for the first time to be $(8.06 \pm 0.31 (\rm{stat}) \pm 0.38 (\rm{syst}))\times 10^{-7}$, approximately twice the non-resonant VMD prediction.

\end{abstract}

\maketitle
An electromagnetic (EM) Dalitz decay~\cite{dalitz} proceeds via conversion of an off-shell photon into a lepton pair~\cite{Landsberg}.  
The $q^2$-dependent transition form factor~(TFF), where $q^2$ is the four-momentum transfer squared and is equal to the square of the di-lepton invariant mass~\cite{Kroll},
provides access to possible deviations from the standard point-like prediction of quantum electrodynamics (QED)~\cite{Landsberg}.
Thus, a TFF is a sensitive tool to probe the inner structure of the involved hadrons. Within the vector meson dominance (VMD) model~\cite{vmd}, the TFF describes the interaction of the photons with hadrons via an intermediate vector meson $V'$ in the time-like region~\cite{Budnev}. These TFFs are important inputs in the calculation of the hadronic light-by-light  contribution of the anomalous magnetic moment of the muon~\cite{hlbl},  $a_{\mu} = (g_{\mu}-2)/2$, which  is  of great interest in precision tests of the Standard Model~\cite{Jegerlehner,Nyffeler,amu1}   and EM Dalitz decay rates~\cite{Landsberg}. 

Most measured EM Dalitz decays of the light unflavored vector mesons $\rho$, $\omega$, and $\phi$~\cite{Hoferichter,Hanhart} are compatible with the VMD assumption~\cite{Landsberg}. However, some of the experimental results, such as $\omega \to \mu^+\mu^- \pi^0$,  indicate large deviations~\cite{Arnaldi, sai} that are difficult to understand with existing theoretical models~\cite{Ananthanarayan}. A similar situation has recently also appeared in the EM Dalitz decays of charmonium vector mesons, such as in recent measurements of the EM Dalitz decays $J/\psi \to e^+e^-P$ ($P=\pi^0, \eta, \eta'$)~\cite{xinkun,vindy}. The measured  branching fractions of $J/\psi \to e^+e^- \eta/\eta'$  agree well with the VMD prediction~\cite{haibo}, which does not include the contributions of intermediate $V'$ resonances. We refer to this as the non-resonant VMD prediction. However, the branching fraction of  $J/\psi \to e^+e^- \pi^0$,  $(7.56 \pm 1.32 \pm 0.50)\times 10^{-7}$~\cite{xinkun}, measured  by \bes for di-electron invariant masses  ($m_{e^+e^-}$) less than $0.4~ \gevcc$~ using 225 million $J/\psi$ events, deviates by more than three standard deviations from the predictions based on the non-resonant VMD prediction~\cite{haibo}, an effective Lagrangian-based analysis~\cite{chen}, Khuri-Treiman (KT) analysis~\cite{KT}, and dispersion theory~\cite{kubis}. The latter three models~\cite{chen, KT, kubis} also take into account the contribution of the intermediate light vector meson $\rho$.  This discrepancy may disappear after considering the resonant contributions from the isospin conserving decay $J/\psi \to \rho \pi^0$ and the isospin violating decays $J/\psi \to \omega/\phi \pi^0$~\cite{pdg}.  However, the well-known $\rho-\omega$ interference, as proposed by Glashow~\cite{glashow}, may also appear in these decays, as discussed in Ref.~\cite{jpstovp}.

An experimental study of $\rho-\omega$ interference is essential for testing charge symmetry violation (CSV)~\cite{McNamee, Thomas, Goldman}, QCD sum rules (QCDSR)~\cite{Shufman, Maltman}, chiral perturbation theory (ChPT)~\cite{Ulrech, Thomas2}, and hidden local symmetry (HLS) models~\cite{Benayoun, Benayoun1}. This interference has been well established in the pionic cross-section $\sigma(e^+e^- \to \pi^+\pi^-)$~\cite{Ignatov, benedikt, babarpipicross-sect}, and impacts the measurements of the isospin-violating $\omega \to \pi^+\pi^-$ decay rate~\cite{Ignatov}, the running QED coupling~\cite{Anastasi}, the $2\pi$ contribution to the hadronic vacuum polarization (HVP) in $a_{\mu}$~\cite{Ignatov, benedikt, babarpipicross-sect}, as well as tests of isospin violation in asymmetric nuclear matter~\cite{nuclephys} and lattice QCD benchmarks~\cite{lQCD}. In particular, the observed $\rho-\omega$ interference provides essential input for constraining isospin-breaking and QED corrections in lattice QCD~\cite{lQCD, lQCD1}, understanding CSV in nuclear forces~\cite{McNamee, Thomas, Goldman}, and modeling vector meson properties in nuclear matter and hot, dense QCD environments such as the quark-gluon plasma~\cite{Rapp}.

An investigation of $\rho-\omega$ interference in a purely EM channel was previously reported via the photoproduction of electron-positron pairs off carbon~\cite{photo1}, beryllium~\cite{photo2}, and hydrogen~\cite{photo3} targets, but the phases of the interference were found to be different from  those corresponding to pionic channels, such as $e^+e^- \to \pi^+\pi^-$~\cite{Ignatov, benedikt, babarpipicross-sect}. 
 This discrepancy may be resolved by investigating the $\rho-\omega$ interference in the intermediate resonances of $J/\psi \to e^+e^-\pi^0$.  The same $J/\psi$ decay also tests various theoretical models~\cite{Landsberg, haibo, chen, KT, kubis}, could improve the accuracy of $J/\psi \to V' \pi^0$ ($V'=\rho,\omega$) decay rates~\cite{pdg}, and probes G-parity violation~\cite{jpstovp}. This Letter reports the first measurement of the $q^2$-dependent TFF and an improved measurement of the branching fraction of the EM Dalitz decay $J/\psi \to e^+e^- \pi^0$ in the full $m_{e^+e^-}$ spectrum using a data sample of $(10087 \pm 44) \times 10^6$ $J/\psi$ events collected by the \bes detector.

\bes is a general purpose detector operating at the $e^+e^-$ BEPCII collider, and is described in more detail in Ref.~\cite{bes3nim}. Monte Carlo (MC) simulated events based on {\sc Geant4}~\cite{geant4} are used to study the detection efficiency, to optimize the event selection criteria, and to investigate potential backgrounds. Time-dependent effects, such as beam-related backgrounds and detector running conditions during the data collection period,  are included in the MC simulation. An inclusive MC sample of 10 billion events of generic $J/\psi$ decays, generated by {\sc EvtGen}~\cite{evtgen} for known decay modes with branching fractions quoted in Ref.~\cite{pdg} and by {\sc LundCharm}~\cite{lundcharm} for unknown decay modes, is used for background studies. The signal MC sample is generated using {\sc EvtGen}~\cite{evtgen} according to a Lorentz-invariant amplitude~\cite{LI}, where the TFF function is directly extracted from the data while taking into account  the contribution of intermediate $V'$ resonances in the decay as described below. Furthermore, another signal MC sample containing only the non-resonant contribution of the $J/\psi \to e^+e^- \pi^0$ decay is also generated with the same model by setting the TFF to unity. This is also referred to as the point-like process and is used to optimize the event selection criteria and study the $m_{e^+e^-}$ dependent TFF.

We select events of interest by requiring two  oppositely charged tracks and at least two good photon candidates. Charged tracks are reconstructed from the ionization signals measured by the main drift chamber (MDC). The point of closest approach of each charged track to the $e^+e^-$ interaction point (IP) is required to be within $\pm 10.0$ cm along the beam direction, which is the symmetry axis of the MDC and within $\pm 1.0$ cm in the plane perpendicular to the beam. Each charged track is required to be within the active region of the MDC $|\cos\theta| < 0.93$, where $\theta$ is the polar angle with respect to the positron beam along the z-axis. The particle identification (PID) algorithm, based on the specific energy loss $dE/dx$ and information from the time-of-flight (TOF) system and the electromagnetic calorimeter (EMC), forms a likelihood $\mathcal{L}(h)$, where $h$ is a  particle hypothesis of  $e^{\pm}$, $ \pi^{\pm}$ or $K^{\pm}$ charged tracks. One of the charged tracks is  identified as $e^-$ and the other as $e^{+}$ by requiring $\mathcal{L}(e^{\pm}) > \mathcal{L}(\pi^{\pm})$ and $\mathcal{L}(e^{\pm}) > \mathcal{L}(K^{\pm})$. We further suppress pion contamination by requiring  $E/p_{e^{\pm}} > 0.8$ $c$ for those charged tracks with momenta larger than $0.25$ \gevc, where $E$ is the energy deposited in the EMC and $p_{e^{\pm}}$ is the momentum measured by the MDC.

Showers are reconstructed from clusters of energy deposited in the EMC crystals. The shower energy of each photon is required to be greater than 25 MeV in the barrel region ($|\cos\theta| < 0.8$) or 50 MeV in the end-cap regions ($0.86 < |\cos\theta| < 0.92$). In order to improve the reconstruction efficiency and energy resolution, the energy deposited in the nearby TOF is taken into account. Showers poorly reconstructed in the transition region between the barrel and end-cap are discarded. In order to suppress the electronic noise and energy deposition unrelated to the events, the photon time as given by the EMC is required to be less than 700 ns after the collision.

A vertex fit is performed to the two charged tracks to ensure that they originate from a common vertex point. In order to improve the $\pi^0$ mass resolution, a four-constraint (4C) kinematic fit imposing energy and momentum conservation is implemented for the selected charged tracks and the two good photon candidates under the hypothesis of $J/\psi \to e^+e^- \gamma \gamma$. If there are more than two good photon candidates in the event, all $\gamma \gamma$ combinations are tested and the pair that gives the lowest value of chi-square of the kinematic fit, $\chi_{4 \rm C}^2$, is selected. To suppress peaking background from $J/\psi \to \gamma \pi^0\pi^0$, where one of the $\pi^0$ candidates decays in its Dalitz decay mode $\pi^0\to\gamma e^+e^-$, the $\chi_{4 \rm C}^2$ is further required to be less than 100. Candidate events are required to have a di-photon invariant mass, $m_{\gamma \gamma}$, within the range $[0.09, 0.18]$ \gevcc.

The MC studies indicate that the background  is dominated by the radiative Bhabha process $e^+e^- \to \gamma e^+e^-$, especially in the high $m_{e^+e^-}$ region, due to its large cross-section. The background contributes a smooth distribution in the $m_{\gamma \gamma}$ spectrum. This non-peaking background is suppressed by requiring the momentum of the selected  $e^{\pm}$ to be less than 1.45 \gevc~ and the energy of the low energy photon ($E_{\gamma_2}$) used to reconstruct the $\pi^0$ candidate to be larger than 0.14 \gev.

At this stage, the peaking background in $m_{\gamma \gamma}$ spectrum is dominated by events from the radiative decay $J/\psi \to \gamma \pi^0$, in which   the radiative photon converts into  an $e^+e^-$ pair in the detector material.  The distance from the vertex point of the $e^+e^-$ pair to the origin in the $x-y$ plane is defined as $\delta_{xy} = \sqrt{R_x^2+R_y^2}$, where $R_x$ and $R_y$ are the coordinates of the reconstructed vertex point along the $x$ and $y$ directions, respectively, as described in Ref.~\cite{gammacon}. In the distribution of $R_y$ versus $R_x$, the signal events from $J/\psi \to e^+e^- \pi^0$ accumulate in the center of the circle, and the background events from $J/\psi \to \gamma \pi^0$ accumulate in the circles with radii of 3.5 cm and 6.5 cm corresponding to the positions of the beam pipe and inner wall of the MDC, respectively, as described in  Fig.~1 of Ref.~\cite{Supplementary}.  A requirement of $\delta_{xy} < 2$ cm is imposed, which eliminates around $98\%$ of the  $\gamma$ conversion background with a loss of about $20\%$ of the signal events. 

The two-photon process~\cite{twophot}, which proceeds via the interaction of two virtual photons emitted by the lepton pair in an $e^+e^-$ collider experiment, can produce an even $C$-parity state of the pseudoscalar meson. In the case of $\pi^0$ production, the corresponding process $e^+e^- \to e^+e^- \pi^0$  has  the same final state as the signal of interest.  It therefore contributes as a peaking background in the $m_{\gamma \gamma}$ spectrum. An independent sample of  $\psi(3770)$ data with an integrated luminosity of $2.93$~fb$^{-1}$~\cite{benedikt}, which is away from $J/\psi$ resonance and is assumed to provide a pure sample of $e^+e^- \to e^+e^- \pi^0$ events, is used  to study this background.  The two-photon background events mainly accumulate in the bands of $\cos\theta(e^+) > 0.8$ or $\cos \theta (e^-) < -0.8$,  away from the signal that accumulates around $\cos\theta(e^+) = \cos\theta(e^-)$,  as described in Ref.~\cite{Fengyun}. This background is suppressed by requiring  $\cos \theta (e^+) < 0.8$ and $\cos\theta(e^-) > -0.8$. After applying this selection criterion, the number of two-photon events in the $J/\psi$ data sample is evaluated as $N_{J/\psi}=f \cdot N_{\psi(3770)}$, where $N_{\psi(3770)}$ is the number of signal events in the $\psi(3770)$ data sample surviving the same selection criteria, and $f=\frac{\mathcal{L}_{J/\psi}}{\mathcal{L}_{\psi(3770)}}\cdot \frac{\sigma_{J/\psi}}{\sigma_{\psi(3770)}}\cdot \frac{\epsilon_{J/\psi}}{\epsilon_{\psi(3770)}} =3.59 \pm 0.02$ is the scale factor between the two data samples $X$ ($X=J/\psi$ and $\psi(3770)$), which takes into account the integrated luminosities $\mathcal{L}_X$, the cross-sections $\sigma_X$ and the efficiencies $\epsilon_X$~\cite{twophot}. The detection efficiency ratio $\frac{\epsilon_{J/\psi}}{\epsilon_{\psi(3770)}}$ and cross-section ratio $\frac{\sigma_{J/\psi}}{\sigma_{\psi(3770)}}$ are evaluated to be  $4.03\pm 0.02$ and $0.878 \pm 0.001$, respectively, with the {\sc EKHARA} generator~\cite{ekhara}. The value of $\epsilon_{\psi(3770)}$ is reduced due to the selection criteria on the $e^{\pm}$ momentum, as described above. 
The yield $N_{\psi(3770)}=27.9 \pm 13.6$ is obtained by performing an extended maximum likelihood (ML) fit to the $m_{\gamma \gamma}$ spectrum of events selected from the $\psi(3770)$ data sample. Therefore, the expected number of two-photon events for the $J/\psi$ data sample is estimated to be $100.0 \pm 48.8$.

\begin{figure}
  \centering
 \includegraphics[width=0.50\textwidth]{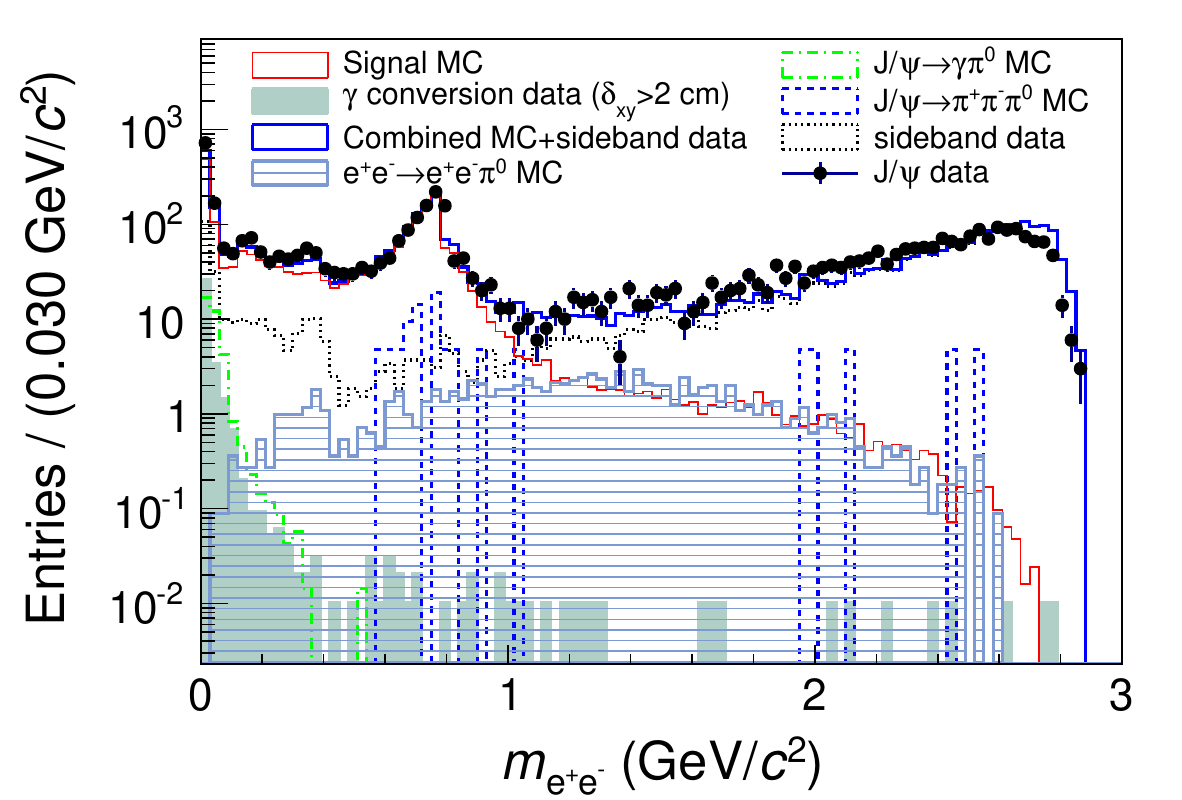}
  
  \caption{ Distribution of $m_{e^+e^-}$ for selected events in data (black dots with error bars), signal MC (red histogram), $J/\psi \to \gamma \pi^0$ MC (green long-dashed dotted histogram), where $\gamma$  converts in material, $\gamma$ conversion data (sky green), $J/\psi \to \pi^+\pi^-\pi^0$ (blue dashed histogram),  sideband data (black dotted histogram), two-photon process MC (grey blue pattern) and combined MC and sideband data (solid blue histogram). The signal MC sample includes  the contributions of both resonant and non-resonant components from $J/\psi \to e^+e^-\pi^0$. The $\gamma$ conversion data were obtained from the events with $\delta_{xy} > 2$ cm~\cite{Supplementary}. }
  \label{mee}
\end{figure}

Figure~\ref{mee} shows the $m_{e^+e^-}$ distribution for surviving events from the data, signal MC, and the remaining background processes. The contribution from the non-peaking background is dominated by the radiative Bhabha process. This background is evaluated using events in the $\pi^0$ sideband regions, defined as [0.02,0.08] and [0.19,0.30] \gevcc. The peaking background is dominated by the decay $J/\psi \to \pi^+\pi^-\pi^0$ ($99.4 \pm 3.8$ events including the $\rho$ resonance and due to the $\pi^\pm$ mis-identified to be $e^\pm$),  $J/\psi \to \gamma \pi^0$ ($35.2 \pm 1.7$ events), $J/\psi \to \gamma \pi^0 \pi^0$ ($2.8 \pm 0.2$ events), and  $J/\psi \to \omega \pi^0$, $\omega \to \pi^+\pi^-$ ($0.5 \pm 0.1$ events), where uncertainties include statistical and systematic uncertainties. The expected numbers of background events are evaluated using large MC samples normalized to their corresponding branching fractions quoted from Ref.~\cite{pdg}. The  simulated events are also used for evaluating the resonant contributions of  $J/\psi \to V' \pi^0$, $V' \to e^+e^-$, considered to be part of the signal mode, and are $690.0 \pm 86.6$ events for $\rho \to e^+e^-$  and $85.5 \pm 9.8$ for $\omega \to e^+e^-$, normalized to the branching fractions in Ref.~\cite{pdg}.

\begin{figure}
\centering
\includegraphics[width=0.5\textwidth]{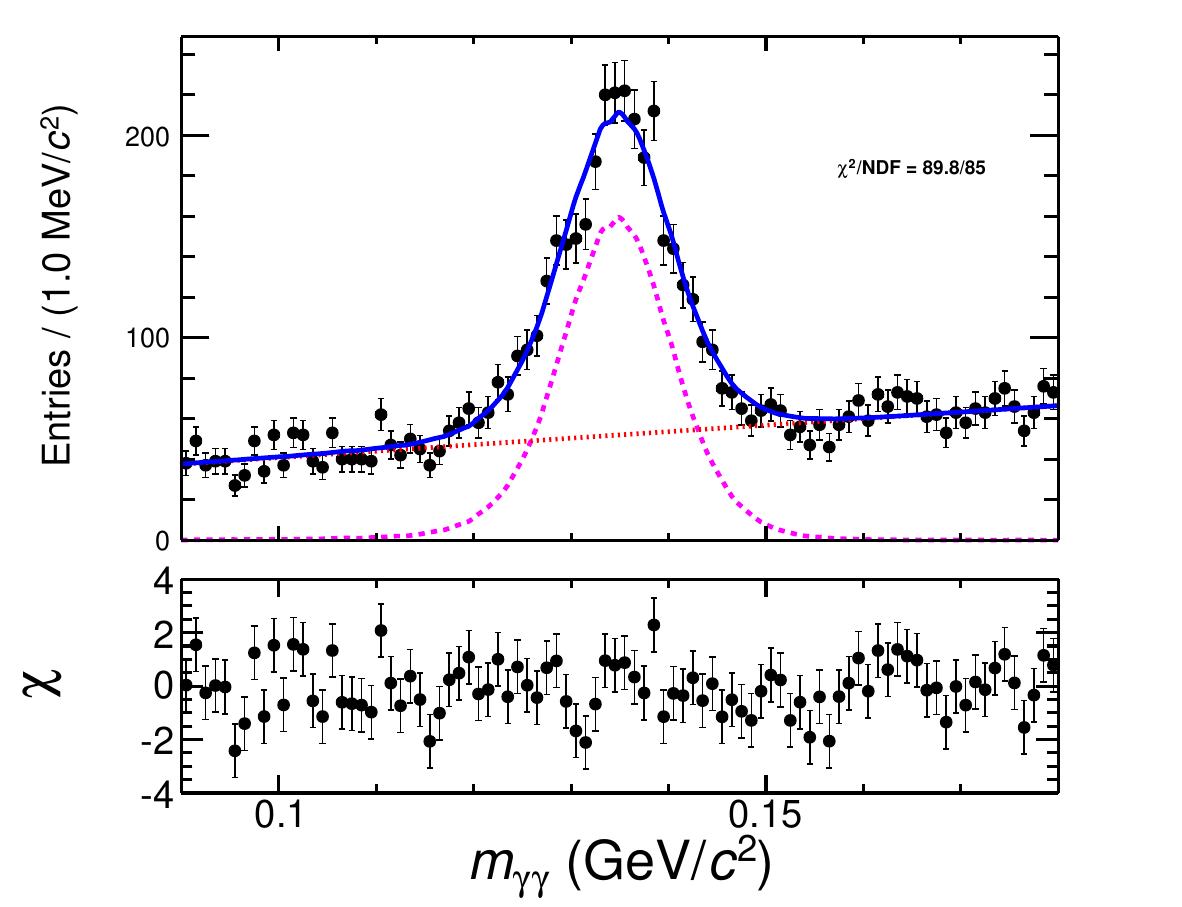}
\caption{The  fit to the $m_{\gamma \gamma}$ distribution (top) together with the normalized fit residuals (bottom). The $p$ value of the fit is  0.34. The black dots with error bars are data, the pink dashed curve the signal, the red dotted curve the  non-peaking background, and the blue solid curve the total projection.}
\label{meeproj}
\end{figure}

The signal yield is determined by performing an extended ML fit to the $m_{\gamma \gamma}$ spectrum of the selected candidate events in data. In the fit, the probability density function (PDF) of the signal is described by the MC simulated shape convolved with a Gaussian function (with free parameters), which compensates for the resolution difference between data and MC simulation. The PDF of the non-peaking background is described by a first-order Chebyshev polynomial function with free parameters. The fit is shown in Fig.~\ref{meeproj} over the full $m_{e^+e^-}$ spectrum, and yields $N_{\rm sig}^{\rm data} = 2347 \pm 66$ events, including both signal and peaking background contributions.  By subtracting the contributions from the  decays $J/\psi \to \pi^+\pi^-\pi^0$, $J/\psi \to \gamma \pi^0$, $J/\psi \to \gamma \pi^0\pi^0$,  $J/\psi \to \omega (\to \pi^+\pi^-) \pi^0$, $J/\psi \to \omega \pi^0\pi^0$,  and the two-photon process $e^+e^- \to e^+e^- \pi^0$,  estimated above, the net signal yield is determined to be $N_{\rm sig} = 2108.9 \pm 82.2$, where the total uncertainties on the peaking backgrounds are added in quadrature with the statistical uncertainty on $N_{\rm sig}$.  The obtained $N_{\rm sig}$ includes the contributions of the non-resonant $J/\psi \to e^+e^-\pi^0$ and the resonant $J/\psi \to \rho/\omega \pi^0$, $\rho/\omega \to e^+e^-$ decays. The branching fraction of $J/\psi \to e^+e^- \pi^0$ is calculated to be $(8.06 \pm 0.31 \pm 0.38) \times 10^{-7}$, where the first uncertainty is statistical and the second is systematic as described below and in the supplemental material~\cite{Supplementary}, by the following formula,
\begin{equation}
\mathcal{B}(J/\psi \to e^+e^- \pi^0) = \frac{N_{\rm sig}}{N_{J/\psi}\cdot \epsilon \cdot \mathcal{B}(\pi^0 \to \gamma \gamma)}.
\label{BFeq} 
\end{equation}
\noindent Here, $\epsilon = 26.3\%$ is the signal selection efficiency obtained from a signal MC sample, depicted in Fig.~\ref{mee},
generated using the fitted TFF function of Eq.~\ref{Eq:tfffull} with a pole mass $\Lambda = 3.686$ \gevcc, as described below to take into account both the resonant and non-resonant contributions of $J/\psi \to e^+e^- \pi^0$. The  $\mathcal{B}(\pi^0 \to \gamma \gamma)$ is the  branching fraction of $\pi^0 \to \gamma \gamma$  from Ref.~\cite{pdg} and  $N_{J/\psi} =  (10087 \pm 44) \times 10^{6}$ is the total number of $J/\psi$ events from Ref.~\cite{cpcjps}.

To avoid the observed $\rho/\omega$ contributions, the branching fraction of $J/\psi \to e^+e^- \pi^0$ for $m_{e^+e^-} < 0.3$ \gevcc~ is determined to be $(4.41 \pm 0.18 \pm 0.21) \times 10^{-7}$, which is consistent with the non-resonant VMD prediction from Ref.~\cite{haibo}, using a similar fit procedure detailed in Ref.~\cite{Supplementary}. The observed deviation from the non-resonant VMD prediction~\cite{haibo}, which was larger in the previous \bes measurement~\cite{xinkun}, disappears with a large $J/\psi$ data sample.  However, our analysis reveals that non-resonant VMD prediction~\cite{haibo} underestimates the total branching fraction by approximately a factor of two, as it does not include contributions from the $\rho/\omega$ resonances.

\begin{figure}
\centering
\includegraphics[width=0.5\textwidth]{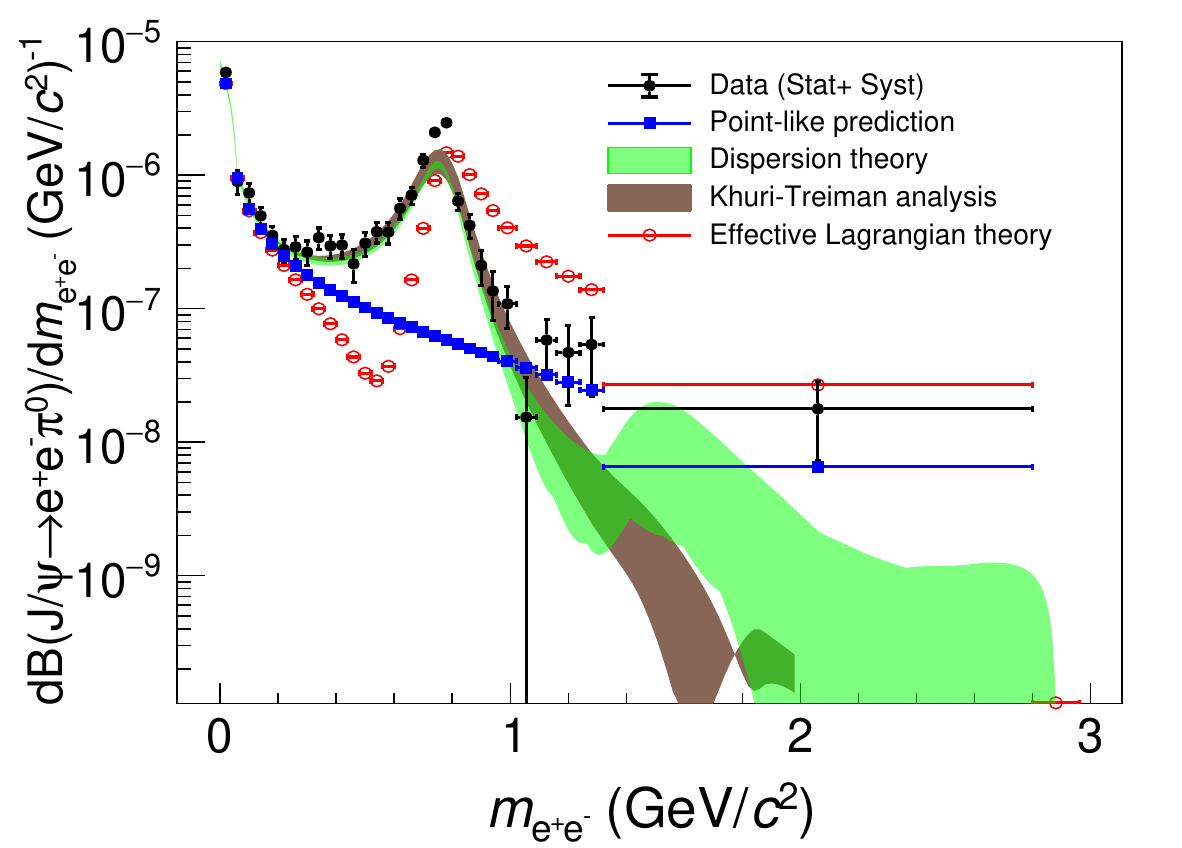}
\caption{The differential branching fraction for  $J/\psi \to e^+e^-\pi^0$  normalized to the bin size versus $m_{e^+e^-}$ for data (black dots with error bars, including the statistical and systematic uncertainties combined in quadrature), the point-like prediction (blue squares with error bars), the effective Lagrangian theory prediction~\cite{chen} (red circles with error bars), KT analysis~\cite{KT} (brown band), and the dispersion theory~\cite{kubis} prediction (green band). }
\label{nsig}
\end{figure}

The $q^2$-dependent TFF of $J/\psi \to e^+e^-$ is studied by comparing the measured differential branching fraction to that of the point-like prediction.
The signal yield in each $m_{e^+e^-}$ bin is  extracted by the same ML fit on the $m_{\gamma\gamma}$ distribution as described above. The corresponding detection efficiency ($\epsilon$) is calculated with a signal MC sample generated under the point-like assumption that the TFF is equal to unity.  In order to  handle background from radiative Bhabha events, the fit is performed with various $m_{e^+e^-}$ bin widths with a total of 30 $m_{e^+e^-}$ points detailed in Table 1 of  Ref.~\cite{Supplementary}.
The obtained distribution of the differential branching fractions, calculated after subtracting the aforementioned peaking backgrounds (except the background from the two-photon process) using the signal yields in each bin of $m_{e^+e^-}$, deviates from the point-like model prediction derived from the formula in Eq.~(3.4) of Ref.~\cite{Landsberg},  as well as from the the effective Lagrangian-based analysis~\cite{chen}, KT analysis~\cite{KT}, and dispersion theory~\cite{kubis}, as shown in Fig.~\ref{nsig}. An interference between the $\rho$ and $\omega$ is observed for the first time in the $J/\psi \to e^+e^-\pi^0$ decay. This behavior differs from that oserved in previous measurements via the photoproduction of electron-positron pairs~\cite{photo1, photo2, photo3}. This difference arises from the isospin dynamics involved in the decays $J/\psi \to \rho \pi^0$ and $J/\psi \to \omega \pi^0$~\cite{jpstovp}.

\begin{figure}
\centering
\includegraphics[width=0.5\textwidth]{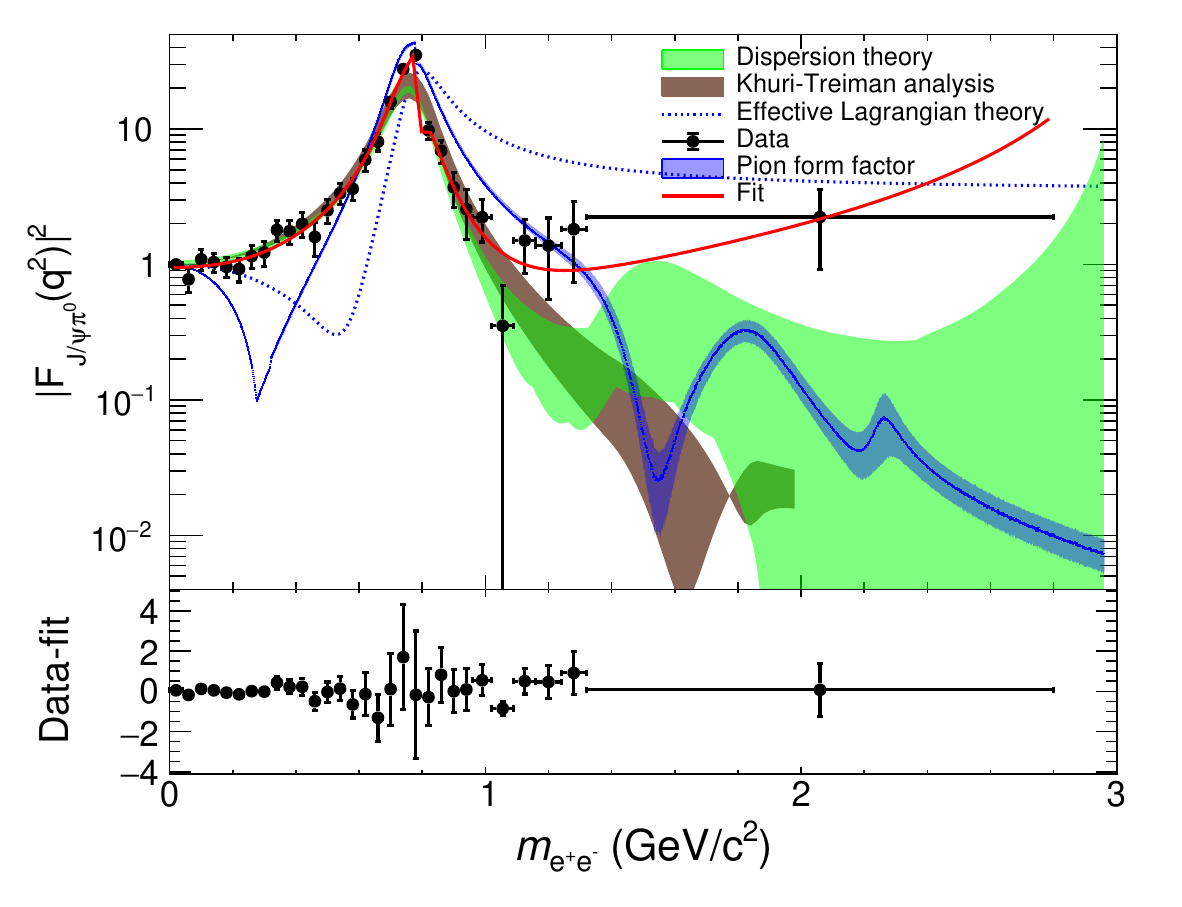}
\caption{(Top) Fit to the TFF versus $m_{e^+e^-}$ in data and (bottom) their fit residual values in each bin. The $p$ value of the fit is 0.82. The black dots with error bars are data, which include both statistical and systematic uncertainties as described in the text and Ref.~\cite{Supplementary}, the blue dotted curve comes from the effective Lagrangian theory~\cite{chen}, the brown and green bands come from the KT analysis~\cite{KT}, and dispersion theory~\cite{kubis}, respectively, blue band is the pion form factor taken from  the BaBar measurement~\cite{babarpipicross-sect} and the solid red curve is the fit result,  which excludes the higher-order $\rho$ resonances.   }
\label{tfffull}
\end{figure}

The TFF ($|F_{J/\psi \pi^0}(q^2)|^2$) in each $m_{e^+e^-}$ bin, shown in Table (1) of Ref.~\cite{Supplementary}, is calculated as the ratio of the corrected signal yield to the prediction of the point-like model integrated over the $m_{e^+e^-}$ bin interval, as shown in Fig.~\ref{tfffull}. To achieve a unity value of the TFF at $q^2=0$, the $|F_{J/\psi \pi^0}(q^2)|^2$ versus $m_{e^+e^-}$ data are normalized using the central value of the TFF from the first $m_{e^+e^-}$ bin.

To study the observed $\rho -\omega$ interference, the TFF is fitted with the following combined function: 

\begin{equation}
F_{J/\psi \pi^0}(q^2)=F_{J/\psi \pi^0}^{\rm GS}(q^2) + A_{\Lambda}\frac{1.}{1-q^2/\Lambda^2},
\label{Eq:tfffull}
\end{equation}

\noindent where $A_{\Lambda}=|A_{\Lambda}| e^{i\phi_{\Lambda}}$ is an amplitude with phase angle $\phi_{\Lambda}$, and $F_{J/\psi \pi^0}^{\rm GS}(q^2)$ includes the contributions of both $\rho$ resonances, which as described by a Gounaris-Sakurai function~\cite{gsmodel}, and $\omega$ resonance, described by a Breit-Wigner function. The $F_{J/\psi \pi^0}^{\rm GS}(q^2)$  is given by the formula in Eq.~(26) of Ref.~\cite{babarpipicross-sect} and has previously used by both the BESIII~\cite{benedikt} and BaBar~\cite{babarpipicross-sect} experiments in the measurement of  the $e^+e^- \to \pi^+\pi^-$ cross-section in the $\rho/\omega$ mass region. The contributions of the $\rho'$, $\rho''$ and $\rho'''$ resonances are excluded in the fit.

Due to limited statistics, the mass and width of  the $\omega$ meson are fixed to their world-average values~\cite{pdg}, while the other parameters are left free during the fit. The obtained fit reveals that the observed $\rho-\omega$ interference pattern in the $J/\psi \to e^+e^-\pi^0$ TFF closely resembles the pattern seen in the  $e^+e^- \to \pi^+\pi^-$ cross-section measurements in the $\rho/\omega$ mass region~\cite{Ignatov, benedikt, babarpipicross-sect}, offering an alternative approach to extract the HVP contribution to $a_{\mu}$. The observed relatively narrow $\rho$ resonance in data arises due to the electron mass, which accounts for the threshold effect in the low-mass region, as well as the interference between resonant and non-resonant processes in $J/\psi \to e^+e^-\pi^0$, as described in Ref.~\cite{Supplementary}. The fit yields the $\Lambda$ value of $3.3 \pm 0.7$ \gevcc, which appears to be consistent with being above the combined mass threshold of $m_{J/\psi\pi^0}$. However, the uncertainty on this fitted value is quite large, limited by statistics  in the high $m_{e^+e^-}$ region. Therefore, the fitted TFF function of Eq.~\ref{Eq:tfffull} with $\Lambda = 3.686$ \gevcc~ is used to generate the signal MC events for efficiency determination, as mentioned above. To properly account for the HVP contribution to $a_{\mu}$, we perform an additional fit including higher order $\rho$ resonances in the mass range $[2m_e, 1.28]$ \gevcc. Details are provided in Ref.~\cite{Supplementary}.

According to Eq.~\ref{BFeq}, the systematic uncertainty  associated with the branching fraction and TFF measurements include those from $N_{\rm sig}$, the number of $J/\psi$ events, reconstruction efficiencies, and secondary branching fractions, details of which are described in Ref.~\cite{Supplementary}. The systematic uncertainty is dominated by the signal model ($0.6\%$), non-peaking background PDF ($2.2\%$), tracking efficiency ($2.4\%$), PID ($1.2\%$), photon reconstruction efficiency ($0.4\%$), and $E/p_{e^{\pm}}$ ($2.4\%$). Total systematic uncertainties for the branching fraction and TFF measurements are determined to be  $4.7\%$ and $6.6\%$, respectively, by adding the individual ones in quadrature.

In summary, using a data sample of $(10087 \pm 44) \times 10^6$ $J/\psi$ events collected by the \bes detector, we measure the TFF of $J/\psi \to e^+e^- \pi^0$  as a function of $m_{e^+e^-}$ for the first time. A clear $\rho-\omega$ interference pattern is observed, which closely matches that of the pionic form factor~\cite{Ignatov, benedikt, babarpipicross-sect}. This result demonstrates the feasibility of probing isospin-violating effects and provides essential input for improving the VMD model~\cite{haibo} and the HLBL contribution to $a_{\mu}$~\cite{Jegerlehner,Nyffeler,amu1}. Furthermore, the observed $\rho-\omega$ interference offers a new way to extract the HVP contribution to $a_{\mu}$~\cite{Ignatov, benedikt, babarpipicross-sect}. After taking into account the contribution of this resonant structure, the branching fraction of $J/\psi \rightarrow e^+e^- \pi^0$ is measured for the first time in the full $m_{e^+e^-}$ spectrum, to be $(8.06 \pm 0.31({\rm stat}) \pm 0.38( {\rm syst}))\times 10^{-7}$, approximately double the non-resonant VMD prediction~\cite{haibo}. These results can be measured with significantly improved precision at future facilities, such as the Super Tau-Charm Factory experiment~\cite{stcf}.


\textbf{Acknowledgement}
The authors thank Bastian Kubis, Sergi Gonzalez-Solis, and Zhi-Hui Guo for providing the data for their theoretical predictions. The BESIII Collaboration thanks the staff of BEPCII and the IHEP computing center for their strong support. This work is supported in part by National Key R\&D Program of China under Contracts Nos. 2020YFA0406300, 2020YFA0406400, 2023YFA1606000; National Natural Science Foundation of China (NSFC) under Contracts Nos. 11635010, 11735014,  11705192, 11935015, 11935016, 11935018, 11950410506, 12025502, 12035009, 12035013, 12061131003, 12192260, 12192261, 12192262, 12192263, 12192264, 12192265, 12221005, 12225509, 12235017, 12361141819; the Chinese Academy of Sciences (CAS) Large-Scale Scientific Facility Program; the CAS Center for Excellence in Particle Physics (CCEPP); Joint Large-Scale Scientific Facility Funds of the NSFC and CAS under Contract No. U1832207; 100 Talents Program of CAS; The Institute of Nuclear and Particle Physics (INPAC) and Shanghai Key Laboratory for Particle Physics and Cosmology; German Research Foundation DFG under Contracts Nos. 455635585, FOR5327, GRK 2149; Istituto Nazionale di Fisica Nucleare, Italy; Ministry of Development of Turkey under Contract No. DPT2006K-120470; National Research Foundation of Korea under Contract No. NRF-2022R1A2C1092335; National Science and Technology fund of Mongolia; National Science Research and Innovation Fund (NSRF) via the Program Management Unit for Human Resources \& Institutional Development, Research and Innovation of Thailand under Contract No. B16F640076; Polish National Science Centre under Contract No. 2019/35/O/ST2/02907;  The Chilean National Agency for Research and Development ANID PIA/APOYO AFB230003 and ANID FONDECYT regular 1230987; The Swedish Research Council; U. S. Department of Energy under Contract No. DE-FG02-05ER41374

\end{document}




\begin{center}
\textbf{\boldmath  \Large{Study of the electromagnetic Dalitz decay of $J/\psi \to e^+e^- \pi^0$:\\
                 Supplemental material}}
\end{center}

\section{An  illustration of signal versus gamma conversion background events}
\begin{figure} [!htp]
\centering
\includegraphics[width=0.8\textwidth]{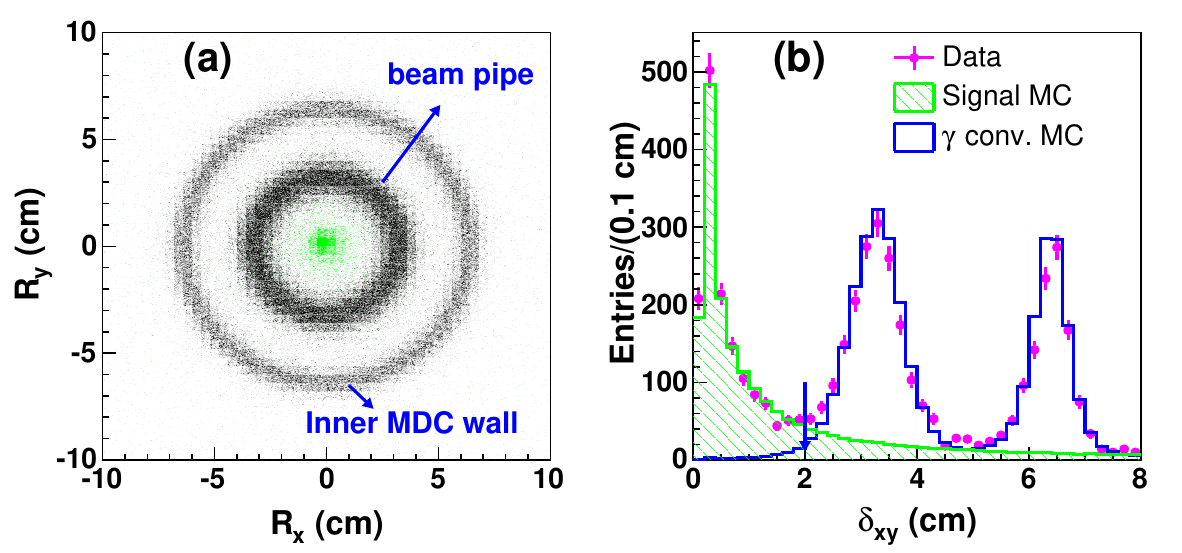}
\caption{ (a) Distribution of $R_y$ versus $R_x$ for the simulated Monte Carlo (MC) events of the $\gamma$ conversion process of $J/\psi \to \gamma \pi^0$ (black dot points) and signal MC events of $J/\psi \to e^+e^- \pi^0$  (green dot points), and (b) the $\delta_{xy}$ distribution of data (pink dot points with error bars), signal MC simulation(green histogram), $\gamma$ conversion MC simulated events (blue histogram). The $\delta_{xy}$ requirement is shown by a solid blue arrow. }
\label{deltaxy}
\end{figure}

To suppress background from $J/\psi \to \gamma \pi^0$, a $\gamma$ conversion finder algorithm~\cite{gammacon} is used. A variable $\delta_{xy} = \sqrt{R_x^2+R_y^2}$, which is the distance from the vertex point of the $e^+e^-$ pair to the origin in the $x-y$ plane, is used to separate the signal from the gamma conversion events. Here $R_x$ and $R_y$ are the coordinates of the reconstructed vertex point along the $x$ and $y$ directions, respectively. A distribution of $R_y$ versus $R_x$ from the MC simulation is shown in Fig.~\ref{deltaxy} (a), where the collected events in the center of the circle are signal events from $J/\psi \to e^+e^- \pi^0$, and the region between the inner and outer hollow circles occurred in the positions of the beam pipe and inner wall of the MDC, respectively, are background  from $J/\psi \to \gamma \pi^0$.
A comparison of the $\delta_{xy}$ distribution from data with the signal and background events from MC simulation is shown in Fig.~\ref{deltaxy} (b).

Prior to this requirement, the yield of $J/\psi \to \gamma \pi^0$ is extracted by fitting the $\delta_{xy}$ distribution in both di-photon invariant mass  ($m_{\gamma \gamma}$) signal ($m_{\gamma \gamma} \in [0.115, 0.155]$ \gevcc) and sideband regions ($m_{\gamma\gamma} \in [0.07, 0.11] \cup [0.16, 0.2]$ \gevcc). The background fraction in $m_{\gamma \gamma}$ signal region is estimated to be 0.5. The corresponding signal yields in the signal and sideband regions are determined to be $2682.5 \pm 57.8$ and $1157.3 \pm 41.1$, respectively. The signal selection efficiency of $0.64\%$ is evaluated using the simulated signal MC sample of $J/\psi \to \gamma \pi^0$ decays. The branching fraction of $J/\psi \to \gamma \pi^0$ is calculated using the net signal events, obtained after subtracting the peaking background contribution from $J/\psi \to \gamma \pi^0\pi^0$ ($23.9 \pm 6.0$ events) and the scaled sideband  events. The measured branching fraction of $J/\psi \to \gamma \pi^0$ agrees with the world average value within $1 \sigma$ uncertainty~\cite{pdg}.  

\section{Non-resonant contribution}
\begin{figure}[!htp]
\centering
\includegraphics[width=3.10in]{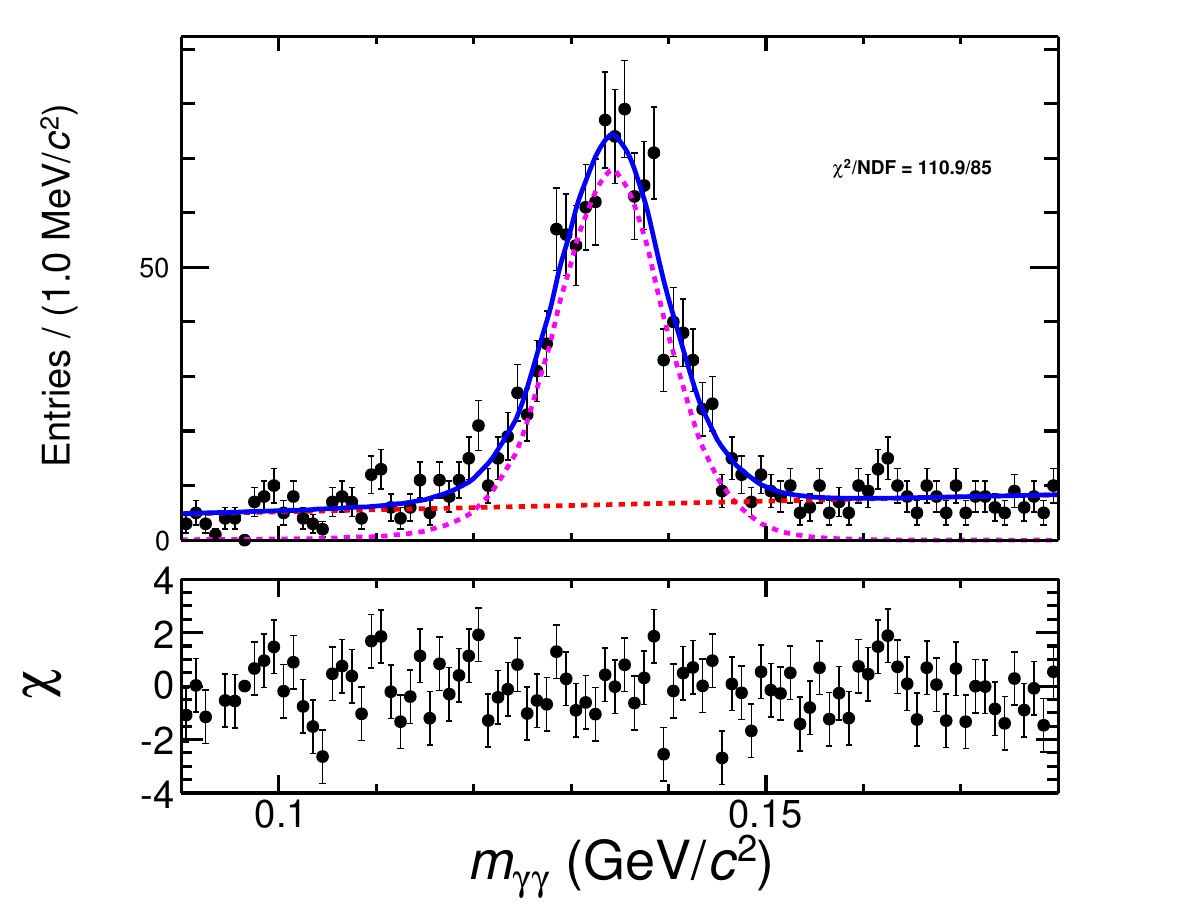}
\caption{Fit to the $m_{\gamma \gamma}$ distribution of $J/\psi$ data for $m_{e^+e^-} < 0.3$ GeV/$c^2$ (top) together with the distribution of normalized fit residuals (bottom). The black dots with error bars represent the data, the pink dashed curve  the signal PDF and  the red dashed curve  the non-peaking PDF. The  solid blue curve is  the total fit result.}
\label{non-reson}
\end{figure}

To calculate the non-resonant contribution of $J/\psi \to e^+e^-\pi^0$, the signal yield is extracted by performing a fit to the $m_{\gamma \gamma}$ distribution for di-electron invariant mass ($m_{e^+e^-}$) less than $0.3$ GeV/$c^2$. The fit yields  $992 \pm 36$  events, which includes both the signal and peaking background contribution from $J/\psi \to \gamma \pi^0$, $J/\psi \to \gamma \pi^0 \pi^0$ and two-photon process of $e^+e^- \to e^+e^- \pi^0$, as seen in Fig.~\ref{non-reson}. Total peaking background in this mass region is predicted to be $84.9 \pm 23.2$ events. After subtracting the peaking background, the net signal yield ($N_{\rm sig }$) is determined to be $907.1 \pm 42.8$ events. The efficiency of the non-resonant contribution of $J/\psi \to e^+e^-\pi^0$ is calculated to be  $20.1\%$ with a signal MC generated with a pole mass of $3.686$ GeV/$c^2$ without including the resonant contribution of $J/\psi \to \rho/\omega\pi^0$. The branching fraction of $J/\psi \to e^+e^-\pi^0$ for $m_{e^+e^-} < 0.3$ GeV/$c^2$ is calculated to be  $(4.41 \pm 0.18 \pm 0.21) \times 10^{-7}$, where the first and second uncertainties are statistical and systematic, respectively.

\section{Systematic uncertainties related to the branching fraction and transition form factor measurements}
The systematic uncertainty associated with the signal modelling is evaluated to be $0.6\%$ by replacing  their shapes from the simulated MC samples by the sum of the two crystal ball (CB) functions~\cite{CB}. The systematic uncertainty of the non-peaking background PDF is evaluated to be $2.2\%$ by replacing the corresponding function with a $2^{nd}$ order Chebyshev polynomial function in the fit. The reliability of the fit is validated by producing a large number of pseudo-experiments containing the same statistics as that of the data. The same fit procedure is performed in each pseudo-experiment, and we consider the relative average difference between the input and output signal yields, which is $0.3\%$, as one of the systematic uncertainties.

A control sample of the radiative Bhabha process $e^+e^- \to \gamma e^+e^-$ is used to explore the efficiencies of tracking and particle identification (PID) for $e^{\pm}$ in the different 2-dimensional bins of momentum versus polar angle. The resulting average differences in efficiency between data and MC are weighted according to the momentum and polar angle of the signal MC, and determined to be $1.2\%$ for tracking and $0.6\%$ for the PID considered for each charged track as systematic uncertainties.  The photon detection efficiency is studied with a control sample of  radiative muon-pair events at $J/\psi$ resonance in which the initial-state-radiation (ISR) photon is predicted using the four-momenta of two charged tracks. The relative difference in efficiency between data and MC is observed to be up to the level of $0.2\%$, and considered to be as systematic uncertainty~\cite{vindy1}. The total systematic uncertainties associated with the tracking, PID and photon reconstruction efficiency are evaluated to be $2.4\%$, $1.2\%$ and $0.4\%$, respectively.

The systematic uncertainty for the $\delta_{xy} < 2$ cm requirement is studied with a control sample of $J/\psi \to \pi^+\pi^-\pi^0$, $\pi^0 \to \gamma e^+e^-$. The signal MC sample for $\pi^0 \to \gamma e^+e^-$ is generated with a simple pole approximation transition form factor (TFF), $F(q^2) = 1+ a_{\pi} q^2/m_{\pi^0}^2$, where $m_{\pi^0}$ is the nominal $\pi^0$ mass and $a_{\pi}= 0.031 \pm 0.004$ is a slope parameter~\cite{pdg}. In order to separate the signal from the background contribution of $\pi^0 \to \gamma \gamma$ in this control sample, the ML fit to the $m_{\gamma \gamma}$ distribution is performed before and after the selection of $\delta_{xy} < 2$ cm  requirement. The corresponding relative difference in efficiencies between data and MC is observed to be $0.12\%$, and taken as the systematic uncertainty. The systematic uncertainty associated with the selection criteria of $E/p_{e^{\pm}}$  is evaluated to be $1.2\%$ for each charged track with a total of $2.4\%$ using the same control sample of  $J/\psi \to \pi^+\pi^-\pi^0$ , $\pi^0 \to \gamma e^+e^-$.  A control sample $J/\psi \to \pi^+\pi^-\pi^0$, $\pi^0 \to \gamma \gamma$ is utilized to study the systematic uncertainty associated with the $4C$ kinematic fit, which is calculated to be $0.4\%$.  The background contribution of $\pi^0 \to \gamma e^+e^-$ in this control sample is eliminated by requiring $|\cos\theta_{\rm heli}| < 0.9$, where $\theta_{heli}$ is the angle between the direction of one of the photons and $J/\psi$ direction in the $\pi^0$ rest frame. The relative difference in efficiency between data and MC, found to be $0.3\%$, is considered as systematic uncertainty.

We vary the requirements of $e^{\pm}$ momentum, $\cos\theta(e^{\pm})$ and the lowest energy of photon used for $\pi^0 \to \gamma \gamma$ reconstruction within one standard deviation of the statistical uncertainties to study the systematic uncertainty of the requirements of these variables. One of the largest values of the relative difference between the signal yields of $J/\psi \to e^+e^- \pi^0$ is $0.4\%$, considered  as the systematic uncertainty.  In the branching fraction measurement of $J/\psi \to e^+e^- \pi^0$ in the full $m_{e^+e^-}$ range, the  simulated MC events, used for the determination of the detection efficiency, are generated with the fit function of TFF measured in this analysis with the $\Lambda$ value of 3.686 GeV/$c^2$.  Two alternative signal MC samples with the simple pole mass $\Lambda$ values of $3.1$ GeV/$c^2$ and $4.0$ GeV/$c^2$ are  generated. One alternative signal MC sample is also generated with the $\rho-\omega$ resonance parameters of dipion channel~\cite{benedikt}. One of the largest relative difference in efficiencies is $0.9\%$, which is considered as systematic uncertainty.  The systematic uncertainty of the $J/\psi$ counting is evaluated to be $0.44\%$ using the inclusive hadronic events of the $J/\psi$ decays. All the possible sources of the systematic uncertainties are summarized in Table~\ref{systfinal}.

\begin{table}[!htp]
\caption{Systematic uncertainties and their sources.}  
\centering
\begin{tabular}{l  c  c }
        \hline
        \hline
Source  & Uncertainty  (for EM Dalitz decay ($\%$)) & uncertainty (for TFF ($\%$)) \\ \hline
\hline

Fixed PDFs                    & 0.55  & 0.55   \\
Fit Bias                      & 0.30  & 0.30         \\
Background modelling          & 2.20  & 2.20          \\
Charged tracks                & 2.40  & 2.40 \\
$e^{\pm}$  PID                & 1.20  & 1.20 \\
Photon detection efficiency   & 0.40  & 0.40 \\
$\chi_{4C}^2$                 & 0.40  & 0.40 \\
Veto of gamma conversion      & 0.12  & 0.12 \\
Photon helicity angle         & 0.30  & 0.30 \\
$\pi^0$ reconstruction        & 1.00  & 1.00 \\
$P_{e^{\pm}}$, $\cos\theta_e^{\pm}$ and $E_{\gamma_2}$  & 0.40  & 0.40 \\
$E/p$                         & 2.60   & 2.60 \\
Form factor                   & 0.90  & ... \\
$\mathcal{B}(J/\psi \to \gamma \pi^0)$ & ... & 4.78 \\
$J/\psi$ counting             & 0.44  & 0.44 \\ \hline
Total                         & 4.67  & 6.62 \\ \hline \hline
\end{tabular}

\label{systfinal}
\end{table}

\section{Di-electron invariant mass dependent transition form factor}

\begin{table}[!htp]
\caption{Background subtracted $N_{\rm sig}^i$, the  measured and quantum electrodynamics (QED) predicted $\mathcal{B}(J/\psi \to e^+e^-\pi)^i$ differential branching fraction (BF) and the transition form factor (TFF )$|F(q^2)|^2$,  for all 44 bins. The first uncertainty is statistical and the second is systematic. Obtained signal yield also includes the total uncertainty of the peaking backgrounds. The QED prediction includes the uncertainty on the branching fraction measurement of $J/\psi \to e^+e^- \pi^0$.  }
  \begin{center}
 \begin{small}
\begin{tabular}{l|l|l|l|l}\hline \hline
  Mass (GeV/$c^2$)     & $N_{sig}$                & BF ($10^{-8}$)      & QED BF ($10^{-8}$)  & TFF  \\\hline
$0.02 \pm 0.02$ & $628.99 \pm 28.88 \pm 39.15$ &  $23.50 \pm 1.10 \pm 1.50$ & $ 19.40 \pm 0.93$ &   $1.21 \pm  0.06 \pm 0.08$ \\
$0.06 \pm 0.02$ & $56.10 \pm 10.30 \pm 4.33$ &  $3.58 \pm 0.66 \pm 0.28$ & $ 3.82 \pm 0.18$ &   $0.94 \pm  0.17 \pm 0.07$ \\
$0.10 \pm 0.02$ & $50.20 \pm 8.25 \pm 3.19$ &  $2.94 \pm 0.48 \pm 0.19$ & $ 2.23 \pm 0.11$ &   $1.32 \pm  0.22 \pm 0.08$ \\ 
$0.14 \pm 0.02$ & $67.50 \pm 9.89 \pm 4.40$ &  $1.98 \pm 0.29 \pm 0.13$ & $ 1.58 \pm 0.08$ &   $1.26 \pm  0.18 \pm 0.08$ \\
$0.18 \pm 0.02$ & $53.80 \pm 8.53 \pm 3.40$ &  $1.41 \pm 0.22 \pm 0.09$ & $ 1.22 \pm 0.06$ &   $1.16 \pm  0.18 \pm 0.07$ \\
$0.22 \pm 0.02$ & $41.60 \pm 7.74 \pm 2.79$ &  $1.11 \pm 0.21 \pm 0.07$ & $ 0.99 \pm 0.05$ &   $1.12 \pm  0.21 \pm 0.08$ \\
$0.26 \pm 0.02$ & $43.30 \pm 7.84 \pm 2.88$ &  $1.16 \pm 0.21 \pm 0.08$ & $ 0.83 \pm 0.04$ &   $1.39 \pm  0.25 \pm 0.09$ \\
$0.30 \pm 0.02$ & $37.80 \pm 7.55 \pm 2.59$ &  $1.06 \pm 0.21 \pm 0.07$ & $ 0.72 \pm 0.03$ &   $1.48 \pm  0.30 \pm 0.10$ \\
$0.34 \pm 0.02$ & $48.90 \pm 8.29 \pm 3.06$ &  $1.36 \pm 0.23 \pm 0.09$ & $ 0.63 \pm 0.03$ &   $2.17 \pm  0.37 \pm 0.14$ \\
$0.38 \pm 0.02$ & $43.30 \pm 7.98 \pm 2.94$ &  $1.18 \pm 0.22 \pm 0.08$ & $ 0.56 \pm 0.03$ &   $2.13 \pm  0.32 \pm 0.14$ \\
$0.42 \pm 0.02$ & $33.10 \pm 6.54 \pm 2.10$ &  $1.20 \pm 0.24 \pm 0.08$ & $ 0.50 \pm 0.02$ &   $2.41 \pm  0.48 \pm 0.15$ \\
$0.46 \pm 0.02$ & $23.20 \pm 6.18 \pm 1.79$ &  $0.87 \pm 0.23 \pm 0.07$ & $ 0.45 \pm 0.02$ &   $1.93 \pm  0.52 \pm 0.15$ \\
$0.50 \pm 0.02$ & $33.90 \pm 6.40 \pm 2.14$ &  $1.24 \pm 0.23 \pm 0.08$ & $ 0.41 \pm 0.02$ &   $3.03 \pm  0.57 \pm 0.19$ \\
$0.54 \pm 0.02$ & $41.00 \pm 6.85 \pm 2.55$ &  $1.51 \pm 0.25 \pm 0.09$ & $ 0.37 \pm 0.02$ &   $4.05 \pm  0.68 \pm 0.25$ \\
$0.58 \pm 0.02$ & $39.10 \pm 6.86 \pm 2.44$ &  $1.50 \pm 0.26 \pm 0.09$ & $ 0.34 \pm 0.02$ &   $4.39 \pm  0.77 \pm 0.27$ \\
$0.62 \pm 0.02$ & $58.50 \pm 8.81 \pm 5.66$ &  $2.26 \pm 0.34 \pm 0.22$ & $ 0.31 \pm 0.02$ &   $7.17 \pm  1.08 \pm 0.70$ \\
$0.66 \pm 0.02$ & $70.20 \pm 9.28 \pm 4.37$ &  $2.83 \pm 0.37 \pm 0.18$ & $ 0.29 \pm 0.01$ &   $9.75 \pm  1.29 \pm 0.61$ \\
$0.70 \pm 0.02$ & $138.00 \pm 13.00 \pm 8.64$ &  $5.17 \pm 0.49 \pm 0.32$ & $ 0.27 \pm 0.01$ &   $19.20 \pm  1.81 \pm 1.20$ \\
$0.74 \pm 0.02$ & $214.00 \pm 15.10 \pm 13.30$ &  $8.36 \pm 0.59 \pm 0.52$ & $ 0.25 \pm 0.01$ &   $33.50 \pm  2.37 \pm 2.08$ \\
$0.78 \pm 0.02$ & $261.00 \pm 17.00 \pm 16.20$ &  $9.88 \pm 0.65 \pm 0.61$ & $ 0.23 \pm 0.01$ &   $42.60 \pm  2.78 \pm 2.64$ \\
$0.82 \pm 0.02$ & $68.40 \pm 8.88 \pm 4.33$ &  $2.55 \pm 0.33 \pm 0.16$ & $ 0.22 \pm 0.01$ &   $11.80 \pm  1.53 \pm 0.75$ \\
$0.86 \pm 0.02$ & $40.70 \pm 7.42 \pm 3.08$ &  $1.68 \pm 0.31 \pm 0.13$ & $ 0.20 \pm 0.01$ &   $8.36 \pm  1.52 \pm 0.63$ \\
$0.90 \pm 0.02$ & $22.10 \pm 6.13 \pm 1.69$ &  $0.84 \pm 0.23 \pm 0.06$ & $ 0.19 \pm 0.01$ &   $4.49 \pm  1.24 \pm 0.34$ \\
$0.94 \pm 0.02$ & $14.40 \pm 5.45 \pm 2.01$ &  $0.54 \pm 0.21 \pm 0.08$ & $ 0.18 \pm 0.01$ &   $3.10 \pm  1.17 \pm 0.43$ \\
$0.99 \pm 0.03$ & $17.00 \pm 5.12 \pm 2.83$ &  $0.65 \pm 0.20 \pm 0.11$ & $ 0.24 \pm 0.01$ &   $2.70 \pm  0.81 \pm 0.45$ \\ 
$1.050 \pm 0.035$ & $2.58 \pm 1.58 \pm 2.02$ &  $0.11 \pm 0.07 \pm 0.08$ & $ 0.25 \pm 0.01$ &   $0.43 \pm  0.26 \pm 0.33$ \\
$1.120 \pm 0.035$ & $10.20 \pm 4.27 \pm 0.84$ &  $0.41 \pm 0.17 \pm 0.03$ & $ 0.22 \pm 0.01$ &   $1.82 \pm  0.76 \pm 0.15$ \\
$1.20 \pm 0.04$ & $9.69 \pm 5.67 \pm 1.20$ &  $0.38 \pm 0.22 \pm 0.05$ & $ 0.23 \pm 0.01$ &   $1.67 \pm  0.98 \pm 0.21$ \\
$1.28 \pm 0.04$ & $9.82 \pm 5.69 \pm 1.23$ &  $0.43 \pm 0.25 \pm 0.05$ & $ 0.20 \pm 0.01$ &   $2.20 \pm  1.28 \pm 0.28$ \\
$2.06 \pm 0.74$ & $59.70 \pm 35.10 \pm 3.85$ &  $2.63 \pm 1.50 \pm 0.17$ & $ 0.97 \pm 0.05$ &   $2.71 \pm  1.60 \pm 0.18$ \\ \hline
Total BF &    & $84.30 \pm 2.57 \pm 1.81$ &   & \\ \hline

\end{tabular}
\end{small}
\end{center}

\label{TFFi}
 \end{table}

\begin{figure}
\centering
\includegraphics[width=0.5\textwidth]{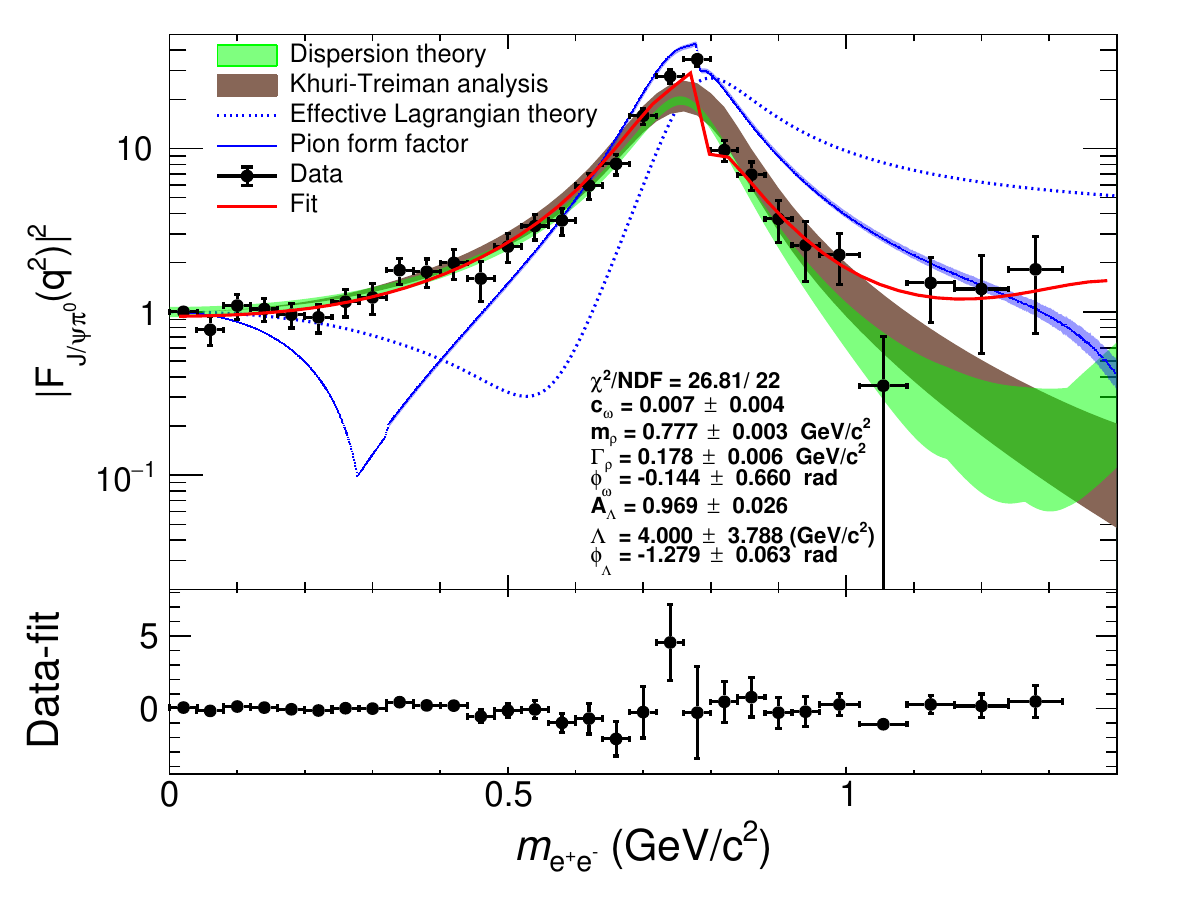}
\caption{(Top) Fit to the TFF versus $m_{e^+e^-}$ in data and (bottom) their fit residual values in each bin. The $p$ value of the fit is 0.45. The black dots with error bars are data, which include both statistical and systematic uncertainties, the blue dotted curve comes from the effective Lagrangian theory~\cite{chen}, the brown and green bands come from the KT analysis~\cite{KT}, and dispersion theory~\cite{kubis}, respectively, brown dashed curve is the pion form factor taken from  the BaBar measurement~\cite{babarpipicross-sect} and the solid red curve is the fit result,  which also includes the higher-order $\rho$ resonances.   }
\label{tfffulllow}
\end{figure}

We fit the TFF as a function $m_{e^+e^-}$ in the mass range of $[2m_e, 1.28]$ \gevcc~ to account for contributions from higher-order $\rho$ resonances. The  Gounaris and Sakurai (GS) function, which describes the $\rho-\omega$ interference pattern, is adopted following previous measurements of the $e^+e^- \to \pi^+\pi^-$ cross-section in the $\rho/\omega$ region by BaBar~\cite{babarpipicross-sect} and BESIII~\cite{benedikt}. The
amplitudes, masses, widths of the  higher $\rho$ states, $\rho(1450)$, $\rho(1700)$, and
$\rho(2150)$ are taken from Ref.~\cite{babarpipicross-sect}. We fix the parameters of $\omega$ according to
the PDG~\cite{pdg} values and ﬂoat the other parameters during the ﬁt.
This fit results are summarized in Fig.~\ref{tfffulllow} and Table~\ref{tab:tfffulllow}.

\begin{table}[!htp]
\caption{Fit parameters and statistical errors of the Gounaris–Sakurai fit of the TFF versus $m_{e^+e^-}$ data, including the higher-order $\rho$ resonances.   }
  \begin{center}
 \begin{small}
\begin{tabular}{l|l}\hline \hline
  Mass (GeV/$c^2$)     & Fitted parameters       \\\hline
  $m_{\rho}$ (GeV/$c^2$ & $0.777 \pm 0.003$  \\
  $\Gamma_{\rho}$ (GeV/$c^2$ & $0.178 \pm 0.003$  \\
  $c_{\omega}$   & $0.007 \pm 0.004$  \\  
  $\phi_{\omega}$   & -$0.144 \pm 0.660$  \\
  $A_{\Lambda}$   & $0.969 \pm 0.026$  \\
  $\Lambda$   & $4.0 \pm 3.788$  \\
  $\phi_{\Lambda}$   & -$1.279 \pm 0.063$  \\
  $\chi^2/{\rm NDF}$   & $26.81/22$  \\  \hline
\end{tabular}
\end{small}
\end{center}

\label{tab:tfffulllow}
 \end{table}

\section{Reason of narrow \boldmath{$\rho$} resonance in data}
\begin{figure}
\centering
\includegraphics[width=0.45\textwidth]{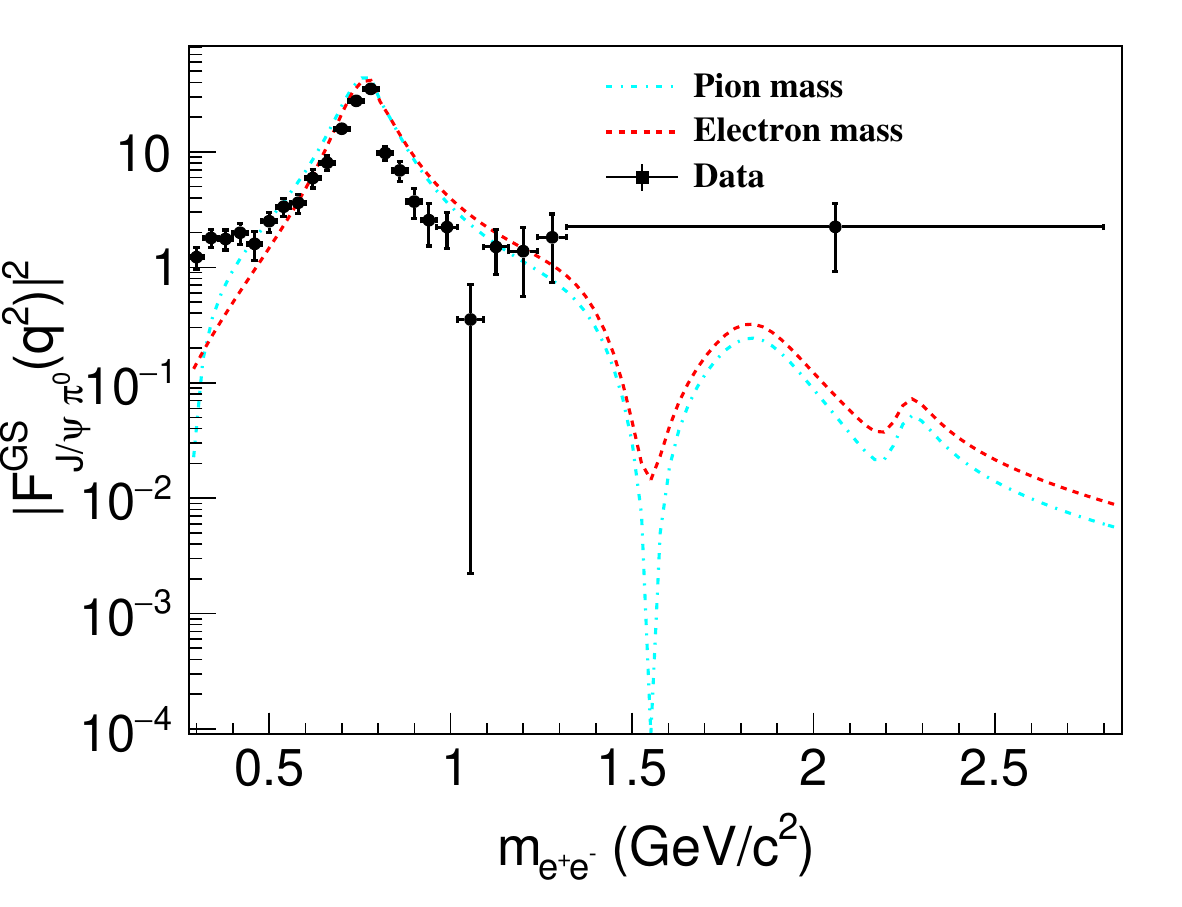}
\includegraphics[width=0.45\textwidth]{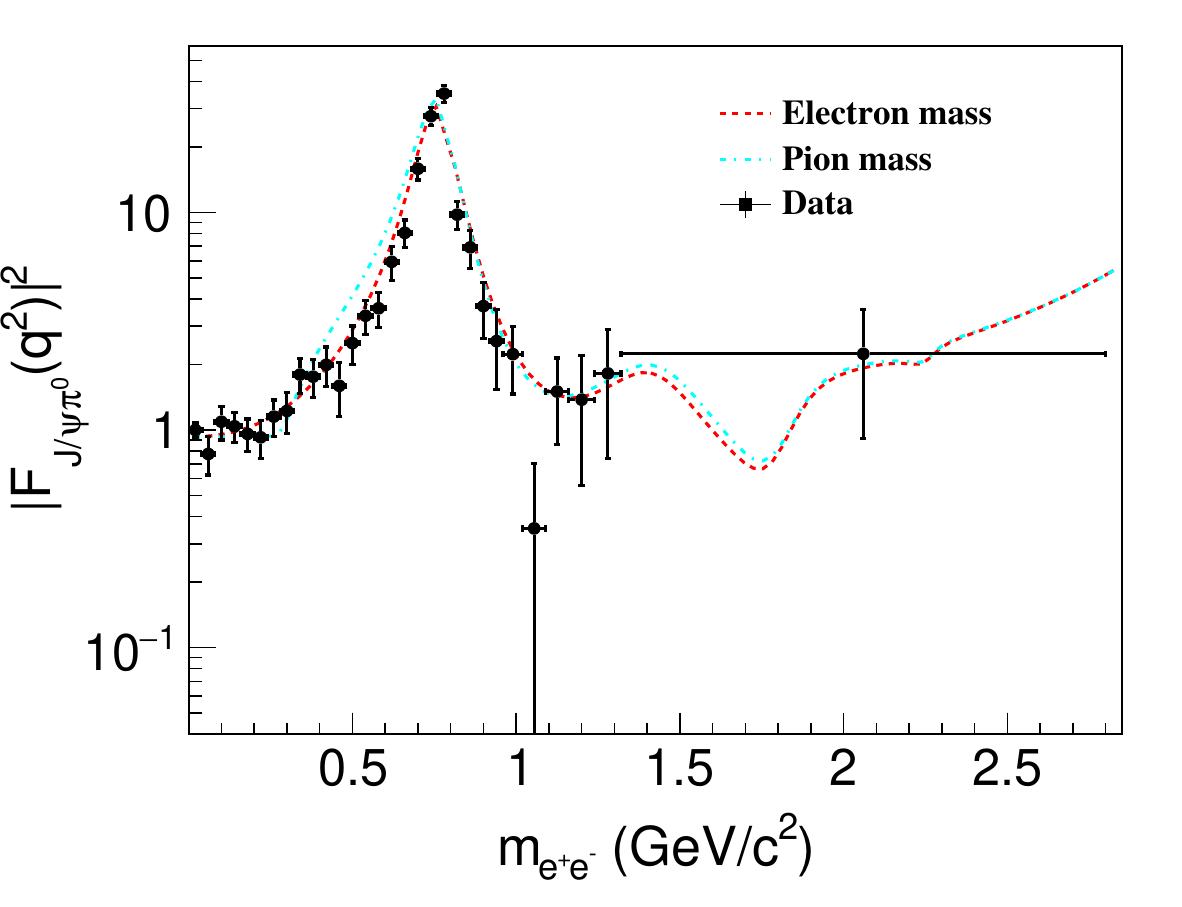}
\caption{The $m_{e^+e^-}$ dependent pion form factor  $F_{J/\psi\pi^0}^{\rm GS}(q^2)$ (left) and $F_{J/\psi\pi^0}(q^2)$ (right) with pion mass (cyan dotted dashed) and electron mass (red dashed) hypotheses, together with  data (black dots with error bars).}
\label{tffdiff}
\end{figure}

To understand the reason for the narrow $\rho$ resonance in data, we plot the $m_{e^+e^-}$ dependent pion form factor $|F_{J/\psi \pi^0}^{\rm GS}(q^2)|^2$ for data and for curves under pion and electron mass hypotheses (Fig.~\ref{tffdiff} (left)), which is described by Eq.~(26) of Ref.~\cite{babarpipicross-sect} with all input parameters taken from Table VI of Ref.~\cite{babarpipicross-sect}. The $|F_{J/\psi \pi^0}^{\rm GS}(q^2)|^2$ curve for the electron mass hypothesis seems slightly different from that of the pion mass hypothesis, and data show a relatively narrow $\rho$ resonance, as already seen in Fig.~\ref{tfffulllow}. However, data are described well by a combined function, $F_{J/\psi\pi^0}(q^2) = F_{J/\psi\pi^0}^{\rm GS}(q^2) + |A_{\Lambda}|e^{i \phi_{\Lambda}} \frac{1}{1-q^2/\Lambda^2}$, described by Eq. (2) in this paper, with fitted parameters $A_{\Lambda}$ and $\phi_{\Lambda}$ from Table~\ref{tab:tfffulllow} and pole mass $\Lambda = 3.686$ GeV/$c^2$ (Fig.~\ref{tffdiff} (right)). The $\rho$ resonance in data appears consistent with this combined function under the electron mass hypothesis. We conclude that the observed relatively narrow $\rho$ resonance in data arises from the electron mass hypothesis and interference between resonant and non-resonant processes in $J/\psi \to e^+e^-\pi^0$.

